# Enhanced bifunctional oxygen catalysis in strained $LaNiO_3$ perovskites

Jonathan R. Petrie[+], Valentino R. Cooper[+], John W. Freeland[±], Tricia L. Meyer[+], Zhiyong Zhang[‡], Daniel A. Lutterman[‡], and Ho Nyung Lee[+*]

[+]*Materials Science and Technology Division, Oak Ridge National Laboratory, Oak Ridge, TN, 37831, USA.*

[±]*Advanced Photon Source, Argonne National Laboratory, Argonne, IL, 60439, USA.*

[‡]*Chemical Sciences Division, Oak Ridge National Laboratory, Oak Ridge, TN, 37831, USA.*

*Supporting Information Placeholder*

**ABSTRACT:** Strain is known to greatly influence low temperature oxygen electrocatalysis on noble metal films, leading to significant enhancements in bifunctional activity essential for fuel cells and metal-air batteries. However, its catalytic impact on transition metal oxide (TMO) thin films, such as perovskites, is not widely understood. Here, we epitaxially strain the conducting perovskite $LaNiO_3$ to systematically determine its influence on both the oxygen reduction (ORR) and oxygen evolution reaction (OER). Uniquely, we found that compressive strain could significantly enhance both reactions, yielding a bifunctional catalyst that surpasses the performance of noble metals such as Pt. We attribute the improved bifunctionality to strain-induced splitting of the $e_g$ orbitals, which can customize orbital asymmetry at the surface. Analogous to strain-induced shifts in the *d*-band center of noble metals relative to Fermi level, such splitting can dramatically affect catalytic activity in this perovskite and other potentially more active oxides.

Advancements in energy storage are essential for driving the development of more sophisticated mobile technologies as well as continuing the trend towards a greener economy. At the forefront of this push are high energy density devices, such as regenerative fuel cells and metal-air batteries.[1] In these and related electrochemical systems, both the oxygen reduction and oxygen evolution reactions (ORR and OER, respectively) are crucial towards successful operation.[2] Traditionally, conductive catalysts incorporating noble metals (e.g., Pt and $IrO_2$) have been used to facilitate these reactions near room temperature.[3] To alleviate costs and poor stabilities during OER in alkaline solutions, significant efforts have focused on transition metal oxides (TMOs) with multivalent Ni, Fe, Co, and Mn, such as $NiFeO_x$.[4] Similar to alloying in noble metals, the majority of research into increasing oxygen activities of TMOs involves cationic doping, which often promotes either the ORR or OER but not bifunctionality.[5] Here, we explore how another factor, i.e. strain, can influence bifunctionality in TMOs.

Contemporary work on high-temperature (>500 °C) oxide catalysis in an aprotic environment (e.g. ORR: $O_2 + 4e^- \rightarrow 2O^{2-}$) has emphasized the importance of tensile strain for activating defects to improve catalytic reactions.[6] In an alkaline environment (e.g. ORR: $O_2 + 2H_2O + 4e^- \rightarrow 4OH^-$), the reaction pathways are different and the activities of these defects are reduced. While recent studies have suggested that tensile strain may increase catalysis in some cobaltites, there is no systematic understanding of strain effects in TMOs.[7] On the other hand, these effects are well-characterized for noble metals, where variations in strain can tailor metal-oxygen (M–O) adsorbate/intermediate bond strengths to optimized levels for catalysis, i.e. the Sabatier principle.[8] For instance, an excessive M–O chemisorption in Pt group metals hinders ORR catalysis. Compressive strain shifts the σ* antibonding states near the Fermi level ($E_F$) lower in energy, decreasing such chemisorption[9]. Among TMOs, the $ABO_3$ perovskites, in which A is traditionally from Groups I-III and B is a transition metal ion with six-fold octahedral coordination, also have electronic structures that are sensitive to strain. Due to strong hybridization between the O $2p$ and the transition metal $d_{z^2}$ and $d_{x^2-y^2}$ lobes in the $BO_6$ octahedra, the σ* states near $E_F$ consist of $e_g$ orbitals.[10,10b] Similar to the Jahn-Teller distortion, strain is known to lift the degeneracy in these symmetry-localized $d_{z^2}$ and $d_{x^2-y^2}$ orbitals, yielding changes in the orbital occupancy, or polarization.[7a, 11] By using strain to control the degree of this $e_g$ orbital splitting and polarization in the octahedra, we anticipated that, analogous to noble metals, we could tailor the perovskite oxygen chemisorption and, hence, catalysis.

To examine the effects of strain on the bifunctional activity of perovskites, we used epitaxial $LaNiO_3$ (LNO) films. LNO is a well-known conducting perovskite at room temperature, eliminating the need for a conducting carbon additive that would influence activity.[12] Moreover, LNO in the bulk form is already more catalytic than Pt towards the OER,[7a] though it has not yet shown higher activity than $NiFeO_x$ and α-$Mn_2O_3$.[4c, 13] Similar to the other oxides, its bifunctionality is limited by a lower activity towards the ORR.[14] Since the rate-determining step (RDS) in both reactions is linked to excessive Ni–O chemisorption to hydroxyl groups, tuning these energies through strain could enhance both ORR and OER.[5b, 9b, 15] Furthermore, the $Ni^{3+}$ ($d^7$) ion in LNO has a stable low-spin (LS) $t_{2g}^6 e_g^1$ configuration.[16] In addition to recent evidence showing that perovskites with an $e_g$-filling of 1 $e^-$ are promising oxygen electrocatalysts[14b], the robust LS $e_g^1$ occupation allows us to focus solely on symmetry changes within the $e_g$ orbitals. Finally, LNO in alkaline solution is stable over a wide range of potentials involving the ORR and OER.[17]

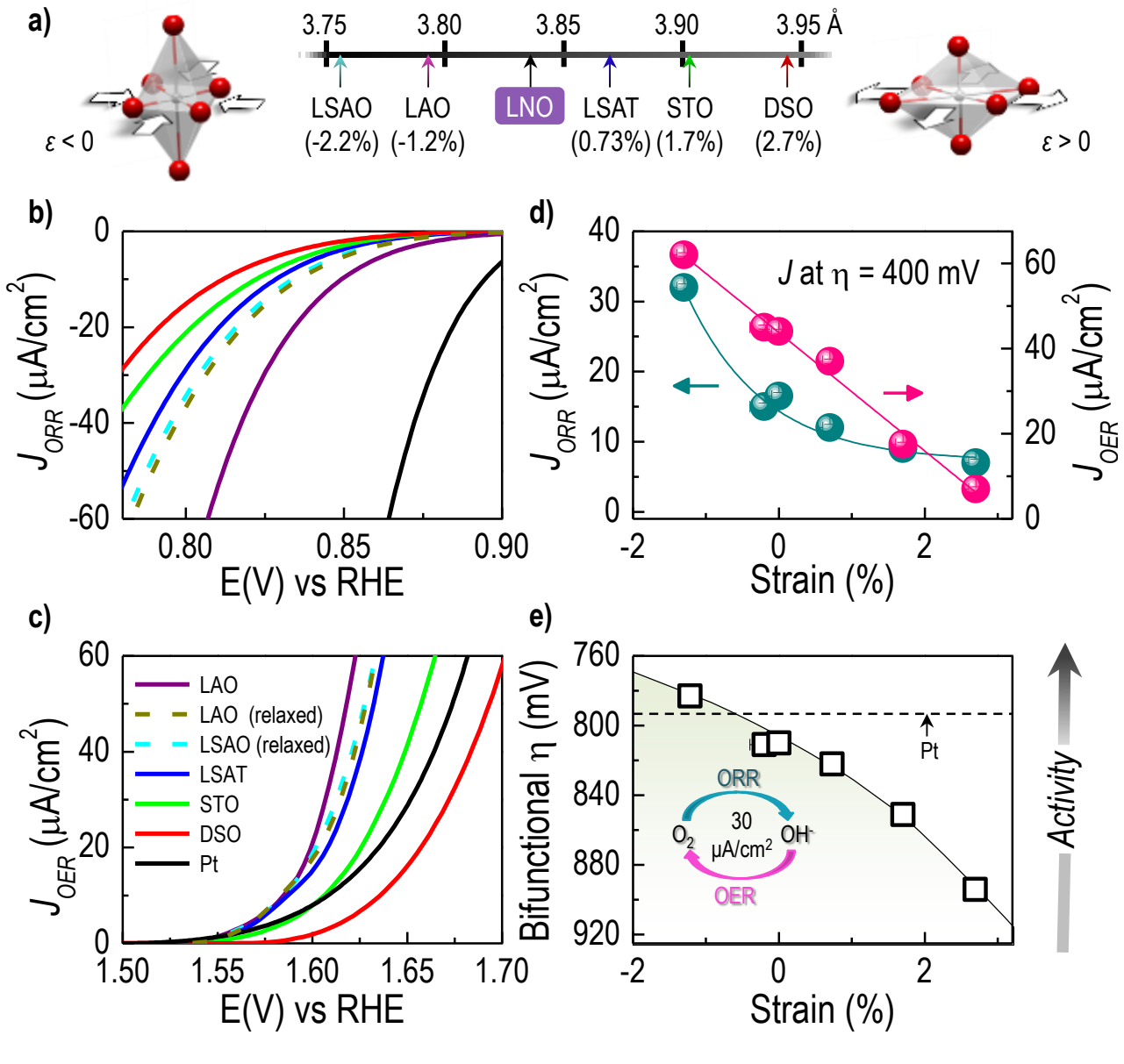

*Figure 1.* Enhanced ORR and OER bifunctional activities using compressive strain. a) Lattice parameters and associated biaxial strain for LNO on various substrates. Polarization curves for the b) ORR and c) OER on these strained LNO films. Strain-relaxed ($\varepsilon \sim 0\%$) films grown on LSAO (10 nm in film thickness) and on LAO (100 nm in film thickness) as well as Pt films are used for comparison. d) The current densities ($J$) of both reactions at overpotentials of η = 400 mV (ORR = 0.823 V and OER = 1.623 V) increase with compressive strain. e) The bifunctional η to attain 30 μA/cm² for both reactions show compressed $LaNiO_3$ surpassing Pt and $IrO_2$.

To systematically introduce strain to (001) LNO, epitaxial films (10 nm in thickness) were deposited by pulsed laser epitaxy (PLE) on a range of lattice-mismatched substrates, which included (001) $LaSrAlO_4$ (LSAO), (001) $LaAlO_3$ (LAO), (001) $(LaAlO_3)_{0.3}(SrAl_{0.5}Ta_{0.5}O_3)_{0.7}$ (LSAT), (001) $SrTiO_3$ (STO), and $(001)_{pc}$ $DyScO_3$ (DSO). As seen by the shift of X-ray diffraction (XRD) peaks in $\theta$-$2\theta$ scans, the near-atomically flat surface from AFM, and electrical transport measurements (Supplementary Figs. S3 and S4), these 10 nm thick (001)-oriented films were all of high quality, conductive, and under biaxial strain from 2.7 to –1.2%. Both pre- and post-test XRD and specific area measurements (see SI) were similar, suggesting minimal instability and phase segregation.[13b, 18] Using XRD reciprocal space mapping (RSM) to determine in-plane lattice parameters, we found all films were coherently strained except for LNO on LSAO, which was relaxed and substantially strain free ($\varepsilon$ = 0 to –0.3%) due to a small critical thickness for strain relaxation. While an additional range of strain states from $\varepsilon$ = –1.2 to 0% was also generated through strain relaxation by increasing the thickness of LNO on LAO from 10 to 100 nm (see Supplementary Fig. S5a), studies on decreasing its thickness on LSAO to minimize strain relaxation are not shown due to systematic limitations measuring films thinner than 5 nm via XRD. As the strain ranged from tensile (+) to compressive (–), the tetragonality, or ratio between out-of-plane to in-plane lattice parameters, increased from 0.96 to 1.04.

We electrochemically characterized the effects of strain on each film in an oxygen-saturated 0.1 M KOH solution (see SI).[13b] A 50 nm Pt film was used for comparison. As depicted in Fig. 1b and c, both the 10 nm films on lattice mismatched substrates as well as the thickness-dependent measurements on LAO (Supplementary Figs. S5b to e) indicate that there is a clear trend towards enhanced catalytic activity with compressive strain for both ORR and OER. Onset potentials clearly shift towards lower overpotentials (η) for both reactions by at least 50 mV between strain of 2.7 to –1.2%. When current densities ($J$) for ORR and OER at a typical operating η = 400 mV are compared (Fig. 1d), we find that the bifunctional activity drastically increases with compressive strain, showing more than an order of magnitude enhancement. Additionally, the Tafel slopes (Supplementary Fig. S6) are similar to Pt for the LNO films at all strain, indicating that the rate-determining steps may be similar.[19] To quantify the bifunctional potential, the total η required to reach 30 μA/cm² ($J_{ORR}$ at η = 400 mV on the $\varepsilon$ = –1.2% LNO) for both ORR and OER are plotted in

Fig. 1e. Due to decreases in η for both reactions, the highest compressively strained LNO (ε = −1.2%) has a bifunctionality exceeding that of Pt. The difference in total η between compressed LNO and Pt over a wider ORR/OER *J* range is shown in Supplementary Fig. S7. Furthermore, as seen in Supplementary Fig. S8, both ORR and OER enhancement does not simply correlate with strain-induced changes in electrical conductivity, precipitating a more in-depth examination of the electronic structure.

As shown in Fig. 2, we used X-ray absorption spectroscopy (XAS) on the Ni-$L_2$ edge to characterize the electronic basis behind the strain-induced modulation in activity.[11b, 20] There is no shift in the polarization-averaged edge, indicating stoichiometric $Ni^{3+}$ with an insignificant amount of oxygen defects in our LNO films (Supplementary Fig. S9). The shoulder features on the leading edges are a consequence of localized carriers in more insulating charge-ordered LNO.[11b] To resolve the orbital splitting energies and $e_g^1$ occupancies between the $d_{z^2}$ and $d_{x^2-y^2}$ orbitals, we employed X-ray linear dichroism (XLD). As seen in Fig. 2a, by detecting the absorption of X-rays polarized both perpendicular (*E* // *c*) and parallel to the film plane (E // *ab*), we were able to probe the respective energies and unoccupied states (holes) of the $d_{z^2}$ and $d_{x^2-y^2}$ orbitals.[10b, 21] Comparing the difference in the peak energy values between the *c*- and *ab*- axis edges, we notice that compressive strain reduces the peak energy of the $d_{z^2}$ orbital compared to the $d_{x^2-y^2}$ orbital while tensile strain has the opposite effect. Furthermore, the $d_{z^2}$ orbital occupancy can be quantitatively determined through sum rules and assumptions originally developed to examine strained LNO heterostructures (see SI).[21-22] By plotting this peak position and orbital occupancy as a function of strain in Fig. 2b, we can see that compressive (tensile) strain-induced orbital splitting results in a lower-energy, high-occupancy $d_{z^2}$ ($d_{x^2-y^2}$) orbital. Although the XLD data shown is from fluorescence yield (FY), the result from total electron yield (TEY) is similar. Note that in both cases, the penetration depth is at least 5-10 nm, which involves probing the entire film.

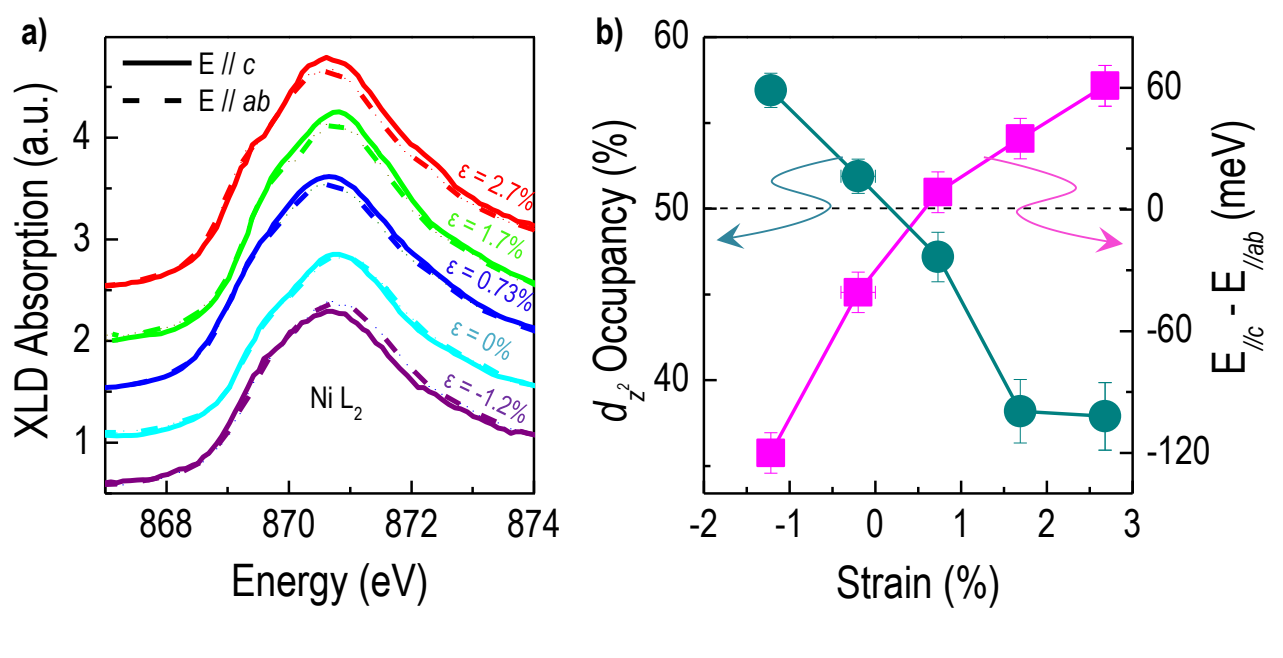

*Figure 2.* XLD characterization of orbital splitting and polarization. a) XLD spectra (FY) of the Ni $L_2$ edge for the strained LNO films. The absorption along the *c*-axis (*ab*-axis) corresponds to holes in the $d_{z^2}$ ($d_{x^2-y^2}$) orbital. b) Using sum rules, strain-induced changes in the occupancy of the $d_{z^2}$ orbital along with the relative energy positions of the $d_{z^2}$ orbital compared to the $d_{x^2-y^2}$ orbital ($E_c - E_{ab}$).

To further investigate the electronic effects at the extreme surface, we computationally modeled the LNO structure using DFT in vacuum for strain from −3 to 3% (see SI). The results are shown in Supplementary Figs. S10 and S11 for both the bulk (5 monolayers) and the surface (first monolayer). Orbital splitting was determined by calculating the centroid of all states for each orbital relative to $E_F$ (analogous to the *d*-band center in metals) and the orbital polarization was found by calculating the orbital $e^-$ occupancy. As seen in Fig. 3a, calculations for the bulk match the experimental XLD results where compressive (tensile) strain favors a more polarized $d_{z^2}$ ($d_{x^2-y^2}$) orbital of lower energy. However, the surface calculations reveal a much different result due to the asymmetry resulting from the lack of an apical oxygen atom on the surface $NiO_6$ octahedra. While the total $e^-$ occupancy of the $e_g$ states is ~1 in all cases, this asymmetry leads to a ~1 eV decrease (increase) in the center of the $d_{z^2}$ ($d_{x^2-y^2}$) orbital relative to $E_F$. Even though strain-induced orbital splitting still occurs at the surface, these offsets result in the $d_{z^2}$ ($d_{x^2-y^2}$) orbital center lying below (above) $E_F$ over the entire strain range. Correspondingly, as shown in Fig. 3b, the orbital polarization at the surface is shifted towards the $d_{z^2}$ orbital; indeed, the occupancy of this orbital in unstrained LNO increases from ~55% of $e_g^1$ in the bulk to ~80% at the surface. Thus, the asymmetry at the surface dramatically favors occupancy of the $d_{z^2}$ orbital over the $d_{x^2-y^2}$ orbital. Similar trends favoring the $d_{z^2}$ orbital occur when the apical position during the RDS is occupied by an adsorbate that is a weaker field ligand on the spectrochemical series, i.e. trending towards vacuum, than the $O^{2-}$ anions in the perovskite $BO_6$ octahedral structure. Here, the hydroxyl ions are indeed weaker field ligands than $O^{2-}$, resulting in a similar asymmetric surface.[23]

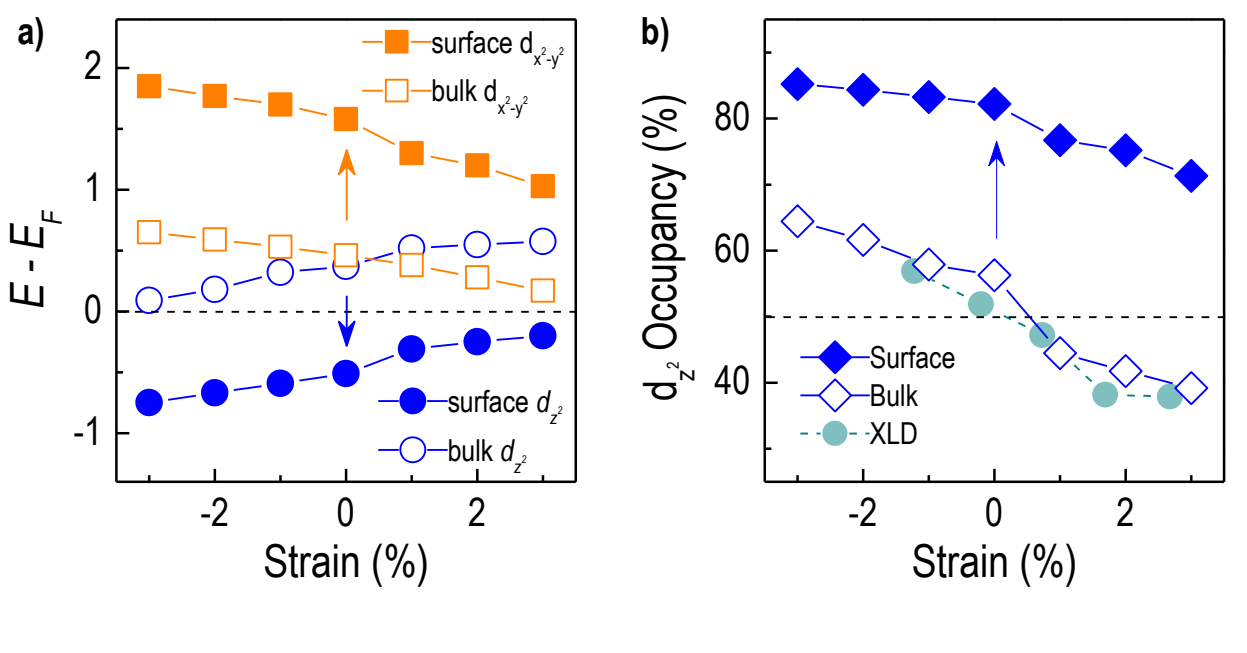

*Figure 3.* Computational modeling of orbital splitting and polarization. a) Position of centroids calculated from density of states for either the $d_{z^2}$ or $d_{x^2-y^2}$ orbitals with respect to $E_F$

as a function of strain for both the LNO bulk (5 monolayers) and the surface (top monolayer). Arrows are used to guide the eye from bulk to surface data. b) The modeled orbital polarization in the bulk closely follows the experimental XLD data (shown from Fig. 2b) while the asymmetric surface occupancy is biased towards $d_{z^2}$.

Interestingly, the trend we detected between strain and oxygen activity in a perovskite such as LNO is comparable to that seen in the Pt group metals. In both cases, a compressive strain-induced weakening of the M–O chemisorption enhances the activity. As seen in Fig. 4b for Pt, variations in bandwidth due to strain within a uniformly-filled *d*-band can shift its center ($E_d$) with respect to $E_F$. Compressive strain broadens the band, essentially lowering $E_d$ to increase σ* occupancy and destabilize the M-O bond. For the small shifts in $E_d$ associated with strain, there is a linear relationship between changes in $E_d$ and M–O chemisorption energy.[23] Extrapolating this concept to LNO, we plotted the combined center of both $e_g$ orbitals ($d_{z^2} + d_{x^2-y^2}$) as a function of strain in Fig. 4a and found no similar monotonic shift that would explain weakening of the M–O bond and subsequent increase in oxygen activity. However, unlike the *d*-band theory for metals, the $e_g$ orbitals at the surface of a correlated oxide become localized, implying, as seen by the DFT results, that (1) strain broadening can be less significant than strain-induced shifts in orbital energies and (2) the $e_g$ orbital with greater $e^-$ occupancy will have a larger influence on M–O chemisorption. Consequently, instead of simply totaling orbital states, we weighted the center of each state to its orbital occupancy at the surface to determine the $e_g$-center ($E_{e_g}$). Due to the disparity of polarization towards the $d_{z^2}$ orbital, its influence in the M–O strength is disproportionate to that of $d_{x^2-y^2}$, which is consistent with a geometric argument postulating greater overlap between O *2p* adsorbate orbitals and the $d_{z^2}$ states at the solution interface. As shown in Fig. 4, $E_{e_g}$ trends towards lower energies with compressive strain, which results in weaker M–O chemisorption and a subsequent enhancement in the catalytic activity. Therefore, by examining both the orbital splitting and polarization at the surface of LNO, we suggest a new parameter, $E_{e_g}$, to predict strain-induced changes in ORR and OER activities in this perovskite and related oxides.

In summary, by investigating the hitherto unexplored effects of strain on the oxygen electrocatalytic activity of LNO perovskite films, we have shown that compressive strain as small as –1.2% can enhance the bifunctional ORR and OER activities above that of the best performing noble metal. While the catalytic trends are akin to strain-induced bandwidth changes in the *d*-band center of noble metals, we have described an $e_g$-center that depends on both orbital splitting and polarization effects at the asymmetric surface to describe shifts in activity. Although applied to a perovskite, there is no reason such strain cannot increase the activity of other TMO-based catalysts, such as $Mn_2O_3$ and $NiFeO_x$. Previously, the possibilities of tensile strain on LNO-based heterostructures have attracted great interest due to physical properties, such as theoretical hints of cuprate-like superconductivity[24]. Our new discovery expands the importance of strain engineering TMOs into the electrochemical realm.

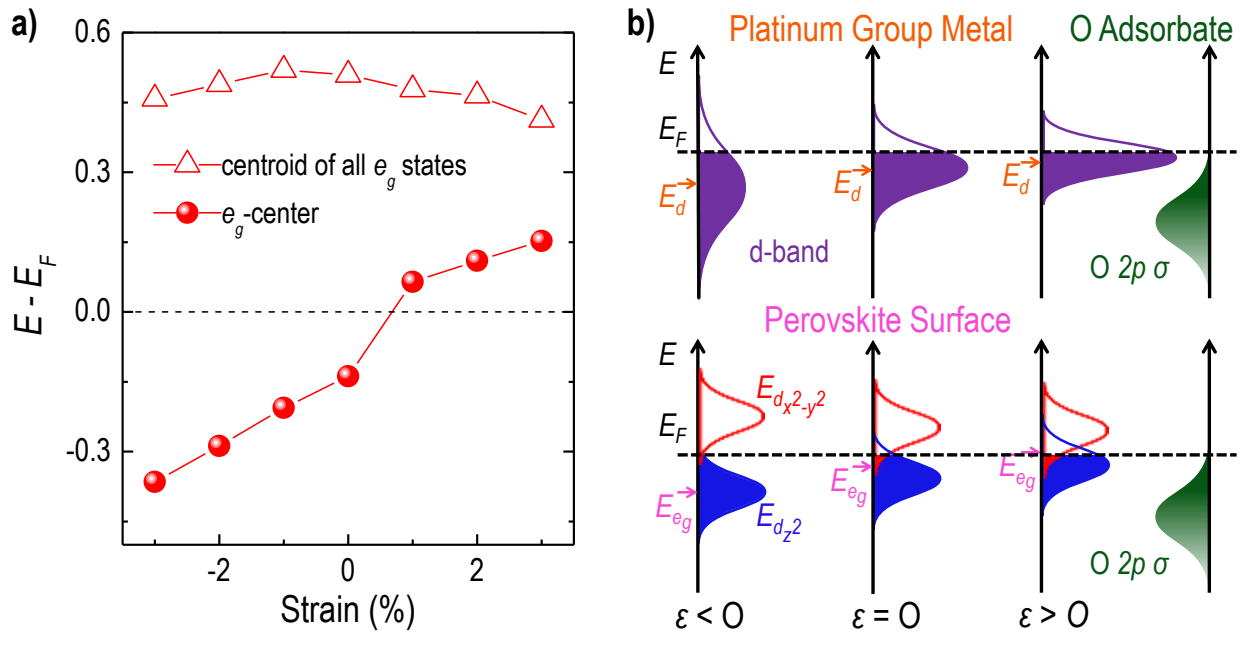

*Figure 4.* $e_g$ center rationalization for strain-induced changes in electrocatalytic activity from dft calculations. a) Position of the centroids calculated from density of states for both all $e_g$ states and the $e_g$-center ($E_{e_g}$) with respect to $E_F$ as a function of strain. The $E_{e_g}$ decreases (increases) with compressive (tensile) strain. b) A schematic relating this trend in perovskites to the *d*-band center ($E_d$) in Pt. While changes in bandwidth play a significant role in $E_d$ shifts for Pt group metals, orbital splitting and polarization have a greater effect at the asymmetric surface of correlated oxides such as LNO.

## ASSOCIATED CONTENT

### Supporting Information

Determination of specific surface area; XRD θ-2θ scans, reciprocal space maps, AFM scan of strained LNO films; electrical transport; ORR/OER activity with strain relaxation through increasing thickness; Tafel curves; comparison of bifunctional overpotential to Pt; ORR/OER activity as a function of electrical resistivity; XAS oxidation state of Ni; DFT modeling. This material is available free of charge via the Internet at http://pubs.acs.org.

## AUTHOR INFORMATION

### Corresponding Author

*hnlee@ornl.gov

## ACKNOWLEDGMENT

We acknowledge S. Okamoto for theoretical insight. This work was supported by the U.S. Department of Energy (DOE), Office of Science (OS), Basic Energy Sciences (BES), Materials Science and Engineering Division (synthesis and physical property characterization) and by the Laboratory

Directed Research and Development Program of Oak Ridge National Laboratory, managed by UT-Battelle, LLC, for the U.S. DOE (electrochemical characterization and theory). Reference electrode preparation was performed as a user project at the Center for Nanophase Materials Sciences, which is sponsored at Oak Ridge National Laboratory by the Scientific User Facilities Division, BES, U.S. DOE. Use of electrochemical testing system was supported by the Fluid Interface Reactions, Structures and Transport (FIRST) Center, an Energy Frontier Research Center funded by the U.S. DOE, OS, BES. Use of Advanced Photon Source was supported by the U. S. DOE, OS, under Contract No.DE-AC02-06CH11357.

# Enhanced bifunctional oxygen catalysis in strained $LaNiO_3$ perovskites

Jonathan R. Petrie[+], Valentino R. Cooper[+], John W. Freeland[±], Tricia L. Meyer[+], Zhiyong Zhang[‡], Daniel A. Lutterman[‡], and Ho Nyung Lee[+*]

[+]*Materials Science and Technology Division, Oak Ridge National Laboratory, Oak Ridge, TN, 37831, USA.*

[±]*Advanced Photon Source, Argonne National Laboratory, Argonne, IL, 60439, USA.*

[‡]*Chemical Sciences Division, Oak Ridge National Laboratory, Oak Ridge, TN, 37831, USA.*
*E-mail: hnlee@ornl.gov

## Thin Film Synthesis

Epitaxial LNO films (10-100 nm in thickness) were grown on various oxide substrates by pulsed laser epitaxy (PLE). The LNO growth temperature, oxygen partial pressure, laser fluence, and repetition rate were optimized at 600ºC, 100 mTorr, 1.5 J/cm$^2$, and 10 Hz, respectively.

## Electrochemical Characterization

The ORR and OER characterization were performed at 25 °C in a 150 ml solution of $O_2$-saturated 0.1 M KOH developed with Sigma-Aldrich KOH pellets and Milli-Q water. Before each test, $O_2$ was bubbled directly into solution while, during electrochemical testing, an over-pressure of $O_2$ was bubbled just over the solution to avoid unnecessary turbulence. A three-electrode rotating disk electrode (RDE) setup was used with a Pt counter electrode and standard calomel (SCE) reference electrode. For conversion to RHE, the reference electrode was calibrated to the reduction of $H_2$ at the counter electrode. Samples were diced into 2.5x2.5 mm$^2$ and the lattice mismatched substrates attached via conductive paste (e.g. silver or carbon-based) to a polished glassy carbon (GC) disc (5 mm in diameter). Conductivity was maintained from the GC to the film by applying the conductive paste along the side of the substrate. All paste was subsequently covered with epoxy to prevent any reaction in solution. Potential was applied via a Biologic SP-200 Potentiostat at 5 mV/s and the samples had a rotating speed of 1600 rpm. Ohmic losses due to the film and solution were determined via a high frequency (~100 kHz) impedance measurement and subtracted from the applied potential to obtain *iR*-corrected currents. Electrochemically-active specific surface areas were determined on the LNO films via double-layer capacitance measurements around the open-circuit potential (OCP), as described in the section later on surface area determination.[1] These measurements produce a systematic surface area used in comparing relative ORR activities; in this case, all specific surface areas were within 10% of the geometric area. Before determining polarization curves, the potential was cycled at least 50 times at a scan rate of 50 mV/s between –0.3 and 0.7 V vs SCE to expose a stable surface under these ORR/OER conditions. Subsequent polarization curves were taken at a scan rate of 5 mV/s at 1600 rpm at least three times to ensure reproducibility. The average of the anodic and cathodic sweeps were taken to minimize capacitive effects. Additional sweeps in Ar-saturated 0.1 M KOH showed negligible activity. To compare activities to noble metals, highly (111) textured Pt thin films film (50 nm in thickness) were respectively sputter deposited or vaporized onto polished glassy carbon discs of the same size

**Structural and Spectroscopic Characterization**

The sample structure was characterized with a high-resolution four circle XRD. In- and out-of-plane lattice constants were determined via reciprocal space mapping (RSM). Temperature-dependent DC transport measurements were conducted using the van der Pauw geometry with a 14 T Physical Property Measurement System (PPMS). Valency and XLD measurements via XAS were performed at the beamline 4-ID-C of the Advanced Photon Source at Argonne National Laboratory.

**Theory**

All density functional theory (DFT) calculations were performed using the Perdew-Burke-Erzenhoff (PBE) exchange-correlation functional and projector augmented wave (PAW) potentials as implemented in the Vienna Ab Initio Simulation Package (VASP v5.3.5). A planewave cutoff of 400 eV and a Hubbard U value of 4 eV on the Ni *d*-states were employed for all calculations. For bulk LNO an 8 × 8 × 2 k-point was employed, yielding optimized lattice parameters of a=5.419 and c=13.010 Å for a structure with an R3cH in good agreement with the experimental values of (a=5.4573, c=13.1462 Å). All surface calculations were performed with a 6 × 6 × 1 k-point mesh. For the surface calculations the in-plane axis was varied from –3 % to +3% strain in steps of 1% and all of the atoms were allowed to fully relax until the forces on each atom was less than 15 meV/Å.

**Determination of surface area**

Even though these samples are thin films, we cannot simply use the geometric are to evaluate the current density; in case there is significant roughness, the specific surface area must be known. To do this, we employ the double-layer capacitance method discussed in McCrory, et.al.[1] where the ECSA (electrochemically active surface area defined as specific area here) is:

$$\mathrm{ECSA} = C_{DL}/C_s = (i_{DL}/\nu)/C_s \qquad (1)$$

where $C_{DL}$ is the double-layer capacitance, $C_s$ is the specific capacitance per unit area on an atomically flat surface, $i_{DL}$ is the double-layer charging/discharging current, and $\nu$ is the scan rate. To determine the area, the $C_s$ for LNO must be known. We used LNO on STO for this since, while all the film appeared near-atomically flat in the AFM, it had the lowest RMS values (see Fig. S3). To ensure this was the case, we double-checked the surface area on a diced 2.5× 2.5 mm$^2$ LNO/STO using a common method incorporating different scan rates (ν) of the reversible $Fe^{3+}/Fe^{2+}$ redox couple found in 0.4 mmol of potassium ferricyanide ($K_3Fe(CN)_6$).[2] At room temperature and in 0.1 M KOH, this produces oxidation/reduction peaks at the surface whose peak current ($I_{peak}$) can be used to determine the specific surface area (A) via manipulation of the Randles-Sevcik equation:

$$A = (3.72 \times 10^{-6}) D^{-1/2} C^{-1/2} n^{-3/2} (I_{peak}\, \nu^{-1/2}) \qquad (2)$$

where ν is the scanning rate (0.2 to 0.005 mV/s), D is the diffusion constant of the iron-based analyte (7.6 × 10$^{-6}$ cm$^2$/s), C is the bulk concentration of said analyte (~0.0004 mol/cm$^3$), and n is the number of electrons in the redox half-reaction, which was '1' in this case for $Fe^{2+}/Fe^{3+}$. Both the scans and the plot of $I_{peak}$ vs. $\nu^{-1/2}$ are shown in Fig. S1. The specific surface from this method is 0.059±0.004 cm$^2$, which is within the experimental range of error for an atomically flat 2.5 × 2.5 mm$^2$ film (0.0625 cm$^2$). We note that this test was performed on an LNO/STO sample that did not undergo the ORR/OER measurements but had the same geometric, i.e. areal, area of the one that did; using this $Fe^{2+}/Fe^{3+}$ method for all the samples would have introduced the possibility of Fe contamination at the surface.

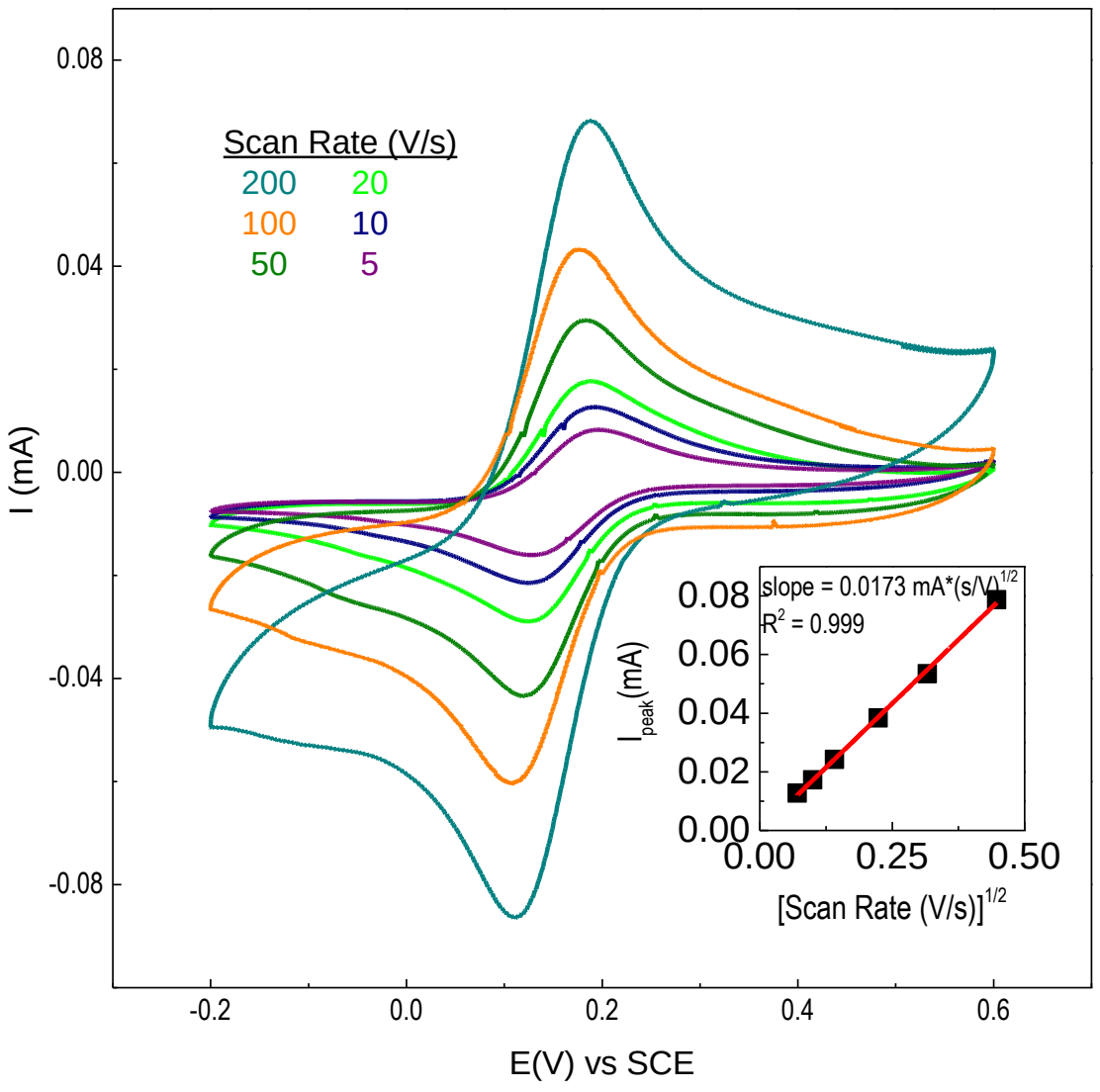

**Figure S1. Effect of changing scan rate on the $Fe^{2+}/Fe^{3+}$ reversible redox peaks.** Cyclic voltammograms of $K_3Fe(CN)_6$ in 0.1 M KOH at scanning rates from 0.2 to 0.005 V/s. The inset shows a plot of the average of the absolute value oxidation/reduction current peaks vs. scan rate to determine surface area.

Once the LNO on STO was independently verified as near-atomically flat, this film was scanned around the OCP (open circuit potential) where all current was assumed due to double-layer charging. The scan for LNO on STO can be seen in Fig. S2 for scan rates from 0.2 to 0.005 V/s. As seen in the inset, by plotting the assumed double-layer current vs. ν, the $C_{DL}$ is found to be ~0.058 mF. Since the specific area is ~ 0.0625 cm$^2$, the Cs ~ 0.93 mF/cm$^2$ for LNO in this environment.

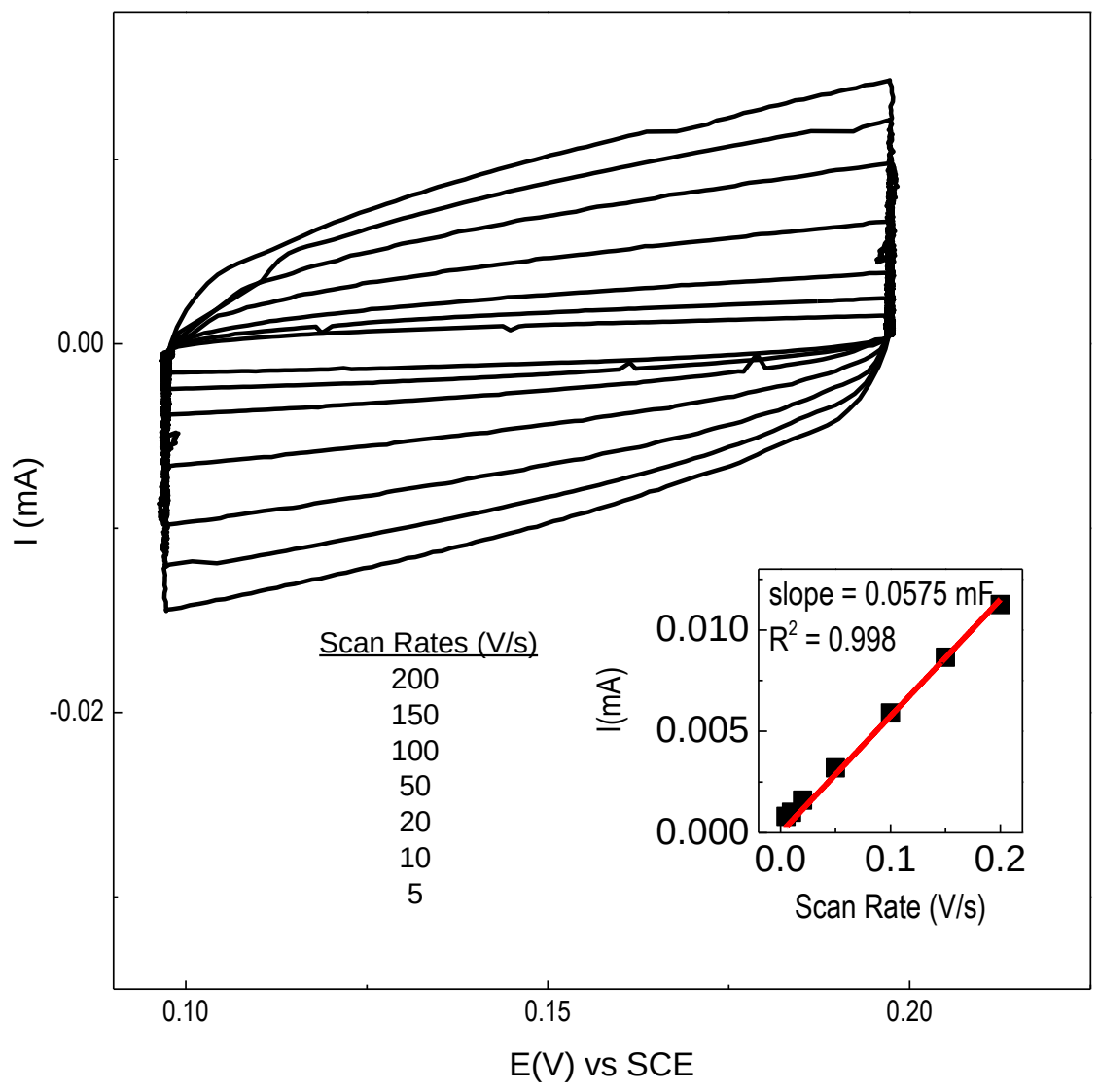

**Figure S2. Double-layer capacitance measurements on LNO/STO.** Cyclic voltammograms around the OCP at scanning rates from 0.2 to 0.005 V/s in 0.1 M KOH. The inset shows a plot of the double-layer current averaged from the absolute values of the anodic/cathodic scans vs. scan rate to determine the double-layer charging capacitance.

Using this value for LNO, the specific surface areas determined via the double-layer capacitance method for LNO on other substrates can be seen in Table S1. The values are all near 0.0625 $cm^2$ both before and after ORR/OER testing, which indicates a stable surface. Geometric areas were calculated from a digital photograph of each sample. The specific areas were all within 10% of the geometric are and the roughness factor (specific area/geometric area) of all these are ~1; however, due to the dicing process, the geometric area of other substrates are not exactly 2.5 × 2.5 $mm^2$, as seen in the slight disparity of areas in the films. Only post-test surface areas were used for determination of ORR and OER current density. In addition, the surface area for the Pt sputtered onto a glassy carbon disk, as determined via either $H_2$ stripping[2] with $Fe^{2+}/Fe^{3+}$ redox couple verification, is shown.

**Table S1. Specific areas of LNO and Pt both before and after testing.** Specific areas of LNO on different substrates using the double-layer technique and $C_s$ from LNO on STO. The specific area is close to the geometric area for all films both before and after testing, indicating (1) the inherent flatness of the films and (2) their stability. The sputtered Pt used as a comparison for ORR/OER activity is also included.

| Material/Substrate | Geometric Area | Pre-test specific area | Post-test specific area |
|---|---|---|---|
| LNO/LAO | 0.070±0.001 $cm^2$ | 0.071±0.005 $cm^2$ | 0.069±0.005 $cm^2$ |
| LNO/LSAO | 0.070±0.001 $cm^2$ | 0.072±0.005 $cm^2$ | 0.075±0.005 $cm^2$ |
| LNO/LSAT | 0.064±0.001 $cm^2$ | 0.060±0.005 $cm^2$ | 0.063±0.005 $cm^2$ |
| LNO/STO | 0.063±0.001 $cm^2$ | 0.063 $cm^{2*}$ | 0.062±0.005 $cm^2$ |
| LNO/DSO | 0.069±0.001 $cm^2$ | 0.075±0.005 $cm^2$ | 0.071±0.005 $cm^2$ |
| Pt/glassy carbon | 0.196±0.001 $cm^2$ | 0.201±0.09 $cm^2$ | 0.205±0.09 $cm^2$ |

*****There is no error here since LNO/STO pre-test used to determine $C_s$ for LNO

## Determination of $d_{z^2}$ occupancy using x-ray linear dichroism (XLD)

As described by Wu, et. al.,[3] the ratio of holes (unoccupied states) in the $e_g$ orbitals is calculated through XLD sum rules as:

$$\beta = \frac{h_{z^2}}{h_{x^2-y^2}} = \frac{3I_c}{4I_{ab}-I_c} \qquad (3)$$

where $h_{z^2}$ is the hole occupancy number of the $d_{z^2}$ orbitals, $h_{x^2-y^2}$ is the hole occupancy number of the $d_{x^2-y^2}$ orbitals, $I_c$ is the X-ray absorption along the out-of-plane direction, and $I_{ab}$ is the X-ray absorption along the in-plane direction.

There are two possible states in each $e_g$ orbital, or $n_{z^2} = 2 - h_{z^2}$ and $n_{x^2-y^2} = 2 - h_{x^2-y^2}$, where $n$

corresponds to the electron occupancy number and $h$ corresponds to the electron occupancy number in each respective orbital. Adding the states of both orbitals, $n_{eg} + h_{eg} = 4$, where $n_{eg}$ is the total electron occupancy number and $h_{eg}$ is the total hole occupancy number. Using the prior assumption that $n_{eg} = 1$, $h_{eg} = 3 = h_{z^2} + h_{x^2-y^2}$ and the fraction $d_{z^2}$ occupancy f($d_{z^2}$) can be determined as:

$$f(\,d_{z^2}) = \frac{n_{z^2}}{n_{eg}} = n_{z^2} = 2 - h_{z^2} = 2 - \frac{3\beta}{1+\beta} \qquad (4)$$

To determine the intensities, we took the average of different Pseudo-Voigt linear combinations of Gaussian and Lorentzian curves to fit the XLD data. The error was determined by the standard deviations of the fits. Errors for intensities were added in quadrature when used for determining orbital occupancy.

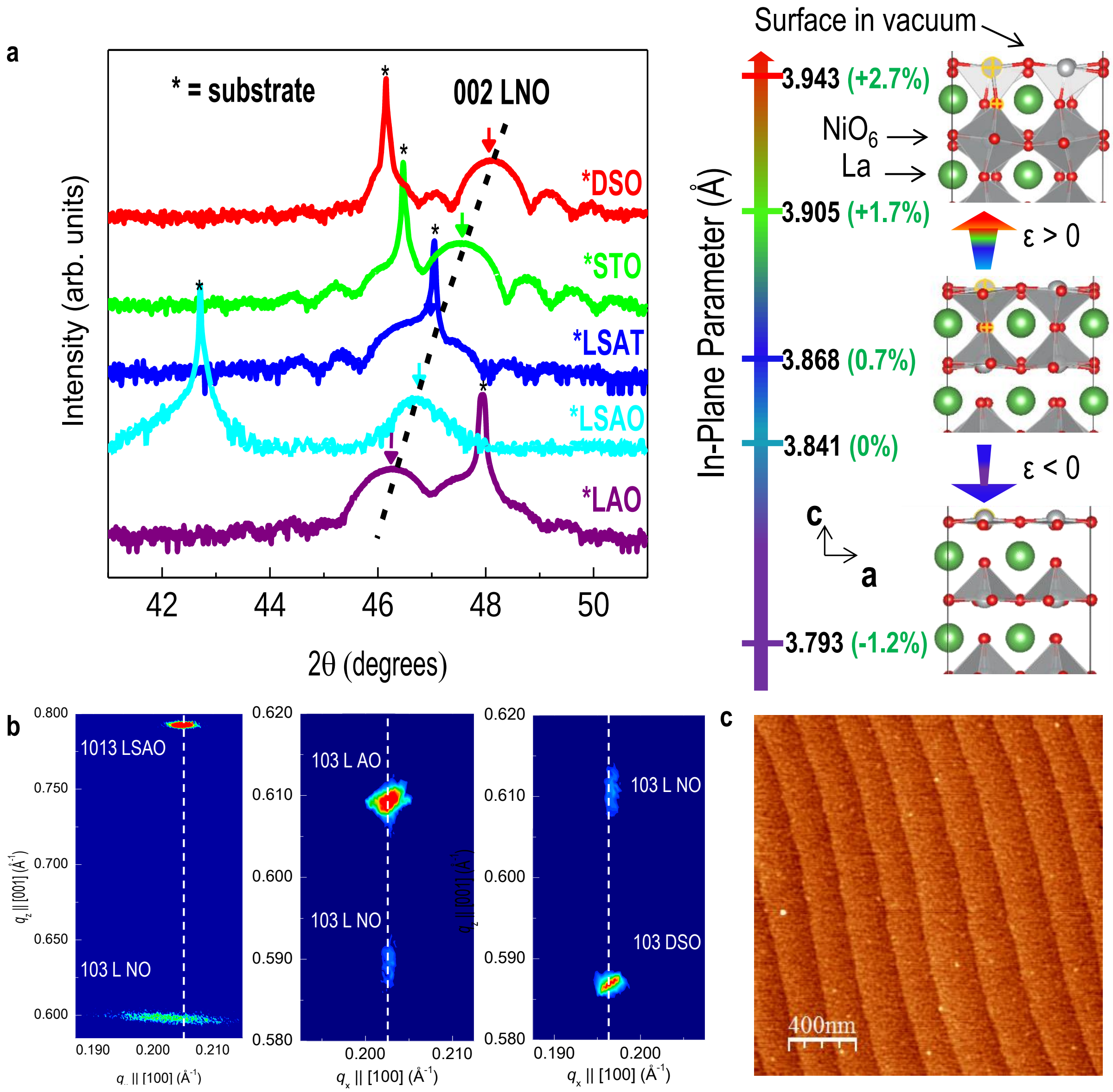

**Figure S3. Structural characterization of strained LNO films. a,** XRD θ-2θ scan of the (002) peak of 10 nm LNO films on DSO, STO, LSAT, LSAO, and LAO substrates in order of increasing compressive strain. The intensity scale is logarithmic. The film on LSAO is almost fully relaxed, leaving LNO in the unstrained state. With increasing compressive strain, there is increased tetragonality, resulting in a raised $Ni^{3+}$ ion in the interrupted octahedral layer at the surface (shown as square pyramidal). XRD scans appeared similar both before and after electrochemical testing. **b,** Reciprocal space mapping of LNO show coherent straining on all substrates except for LSAO, where it is almost fully relaxed ($\varepsilon$ ~ 0%). **c,** Representative atomic force microscopy on an STO substrate, where RMS roughness < 1 nm.

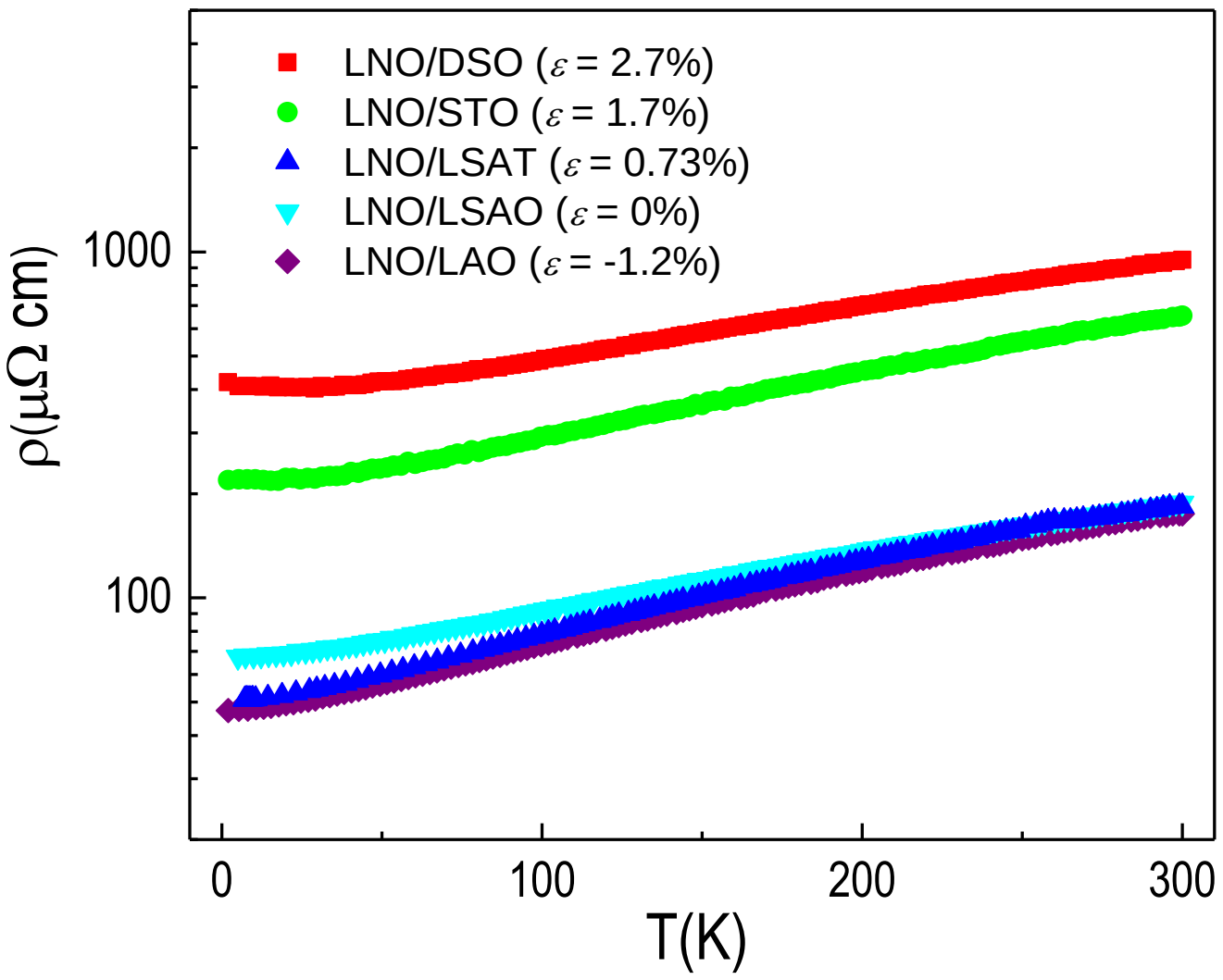

**Figure S4. Electrical characterization of strained LNO films.** Electrical transport data of LNO on different substrates reveal resistivities (ρ) as low as 45 μΩcm as well as residual resistivities ($\rho_{300K}/\rho_{5K}$) of 2-3, reminiscent of the highest quality LNO.[4] Although the electrochemical measurements are already corrected for ohmic losses, the lack of correlation between catalytic activity and resistivity from LSAT to LAO (essentially equivalent DC impedance) argue against the ORR/OER reactions being limited by electron transport through the material.

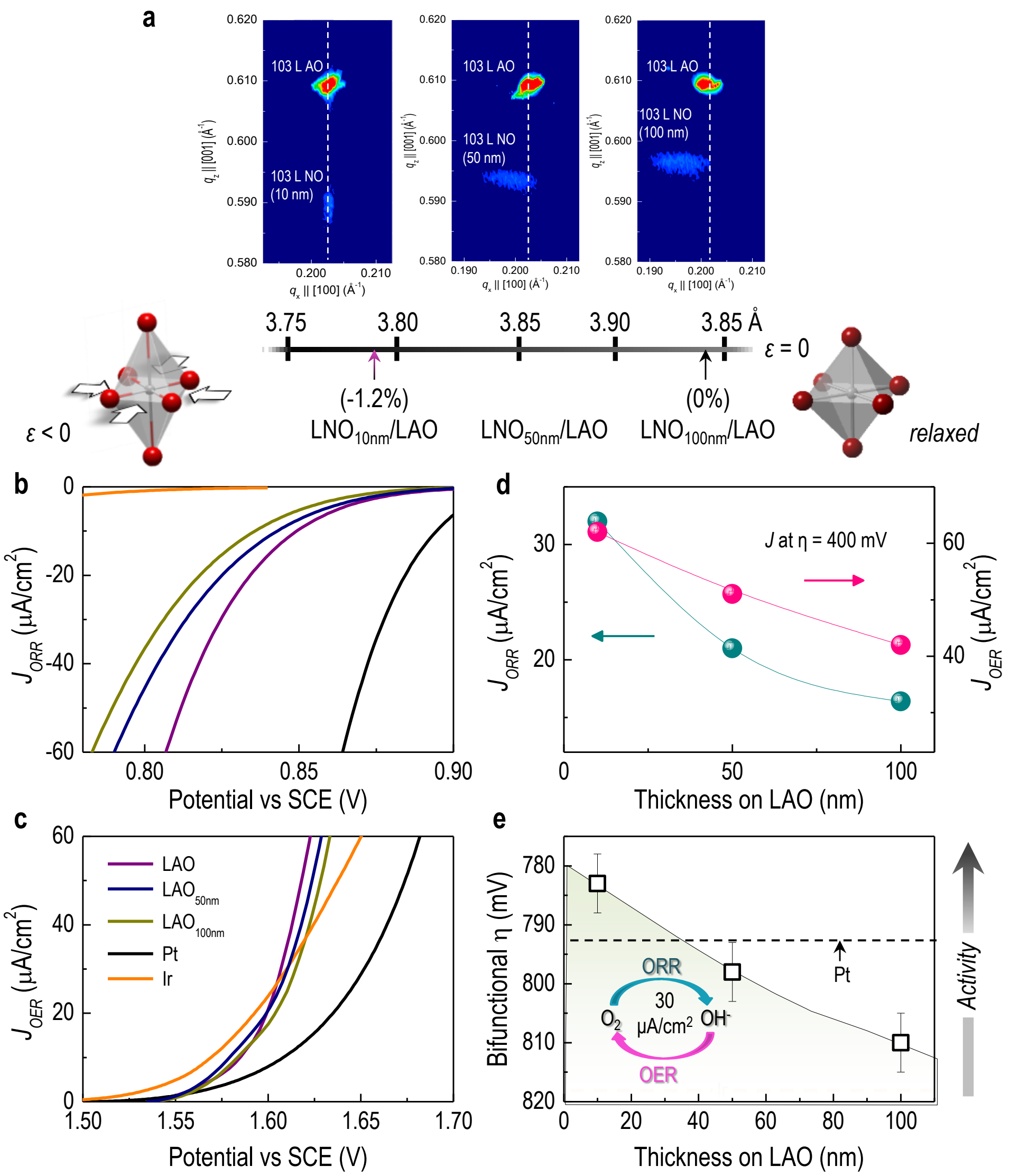

**Figure S5. Relaxation of compressive strain from increasing film thickness reduces both the ORR and OER activity enhancement. a,** Lattice constants and associated biaxial strain for LNO of different thicknesses on LAO. These thickness measurements rely on increased strain relaxation (as seen in the reciprocal space maps around the 103 peaks) with increasing thickness due to a build-up of stress within the film as it becomes thicker. This allows a strain gradient from $\varepsilon \sim -1.2\%$ at 10 nm to $\varepsilon \sim 0\%$ with LNO on the same lattice-mismatched substrate. Electrical conductivity was not significantly different from 10 nm to 100 nm. **b,c,** Polarization curves for the (b) oxygen reduction reaction (ORR) and (c) oxygen evolution reaction (OER) on these LNO films in $O_2$-saturated 0.1 M KOH at a 5 mV/s scan rate and 1600 rpm. The polarization curves for the relaxed 100 nm LNO film on LAO are similar to the relaxed 10 nm LNO film on LSAO. **d,** The current densities ($J$) of both reactions at overpotentials of $\eta$ = 400 mV (ORR = 0.823 V and OER = 1.623 V) increase with compressive strain. **e,** The combined $\eta$ to attain 30 μA/cm$^2$ for both reactions show that the –1.2% compressed $LaNiO_3$ can surpass Pt as a bifunctional catalyst. These thickness tests on a similar substrate further verify the catalytic effects of strain on LNO.

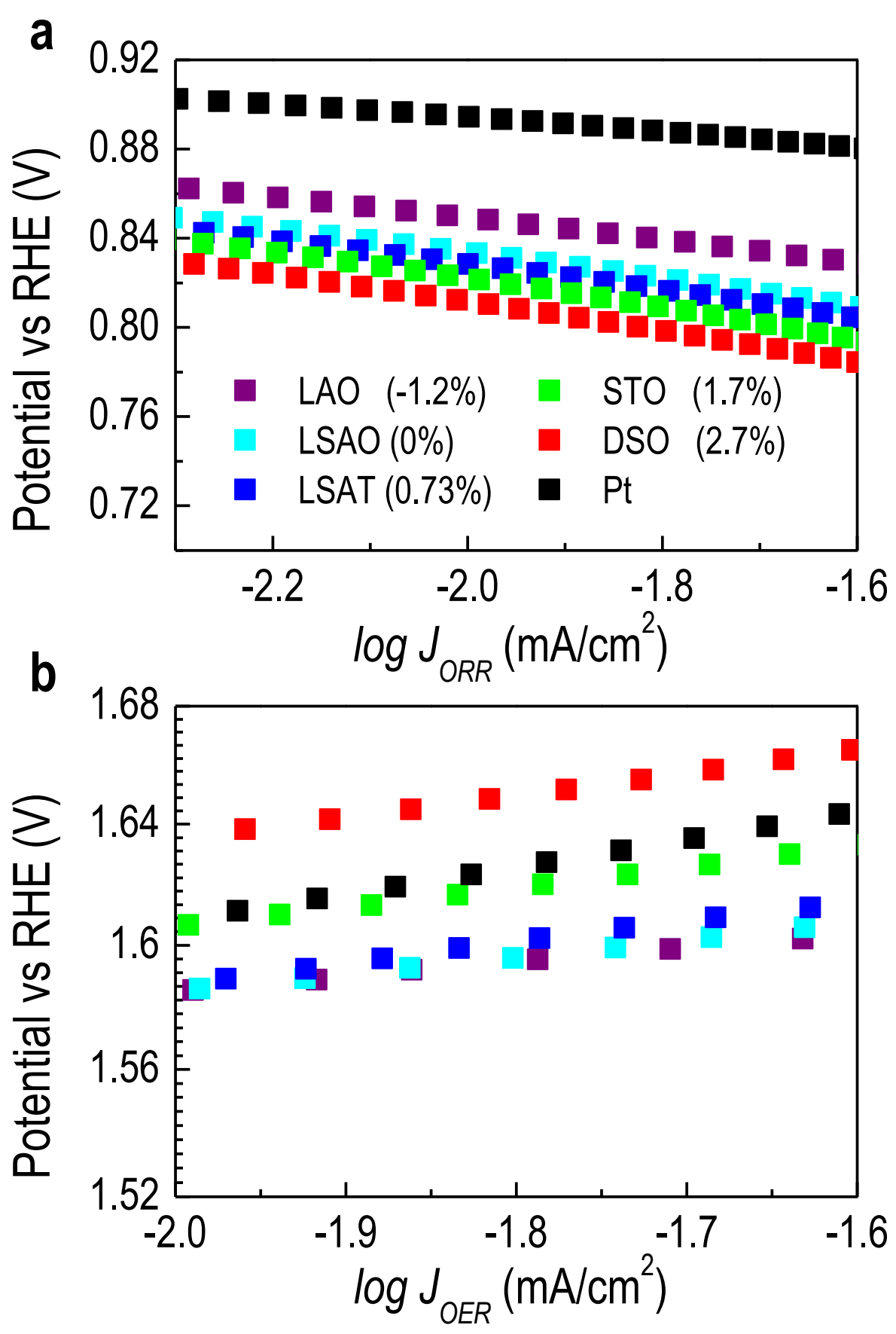

**Figure S6. Tafel curves for ORR and OER current densities around η = 400 mV on LNO films in $O_2$-saturated 0.1 M KOH. a,** All ORR slopes are approximately –60 mV/decade, similar to the Pt curve (■), suggesting a similar number of electrons and similar rate-determining step for all films.[2] **b,** All slopes for OER are approximately 45 mV/decade, suggesting a similar increase in the rate-determining step for this reaction as well.

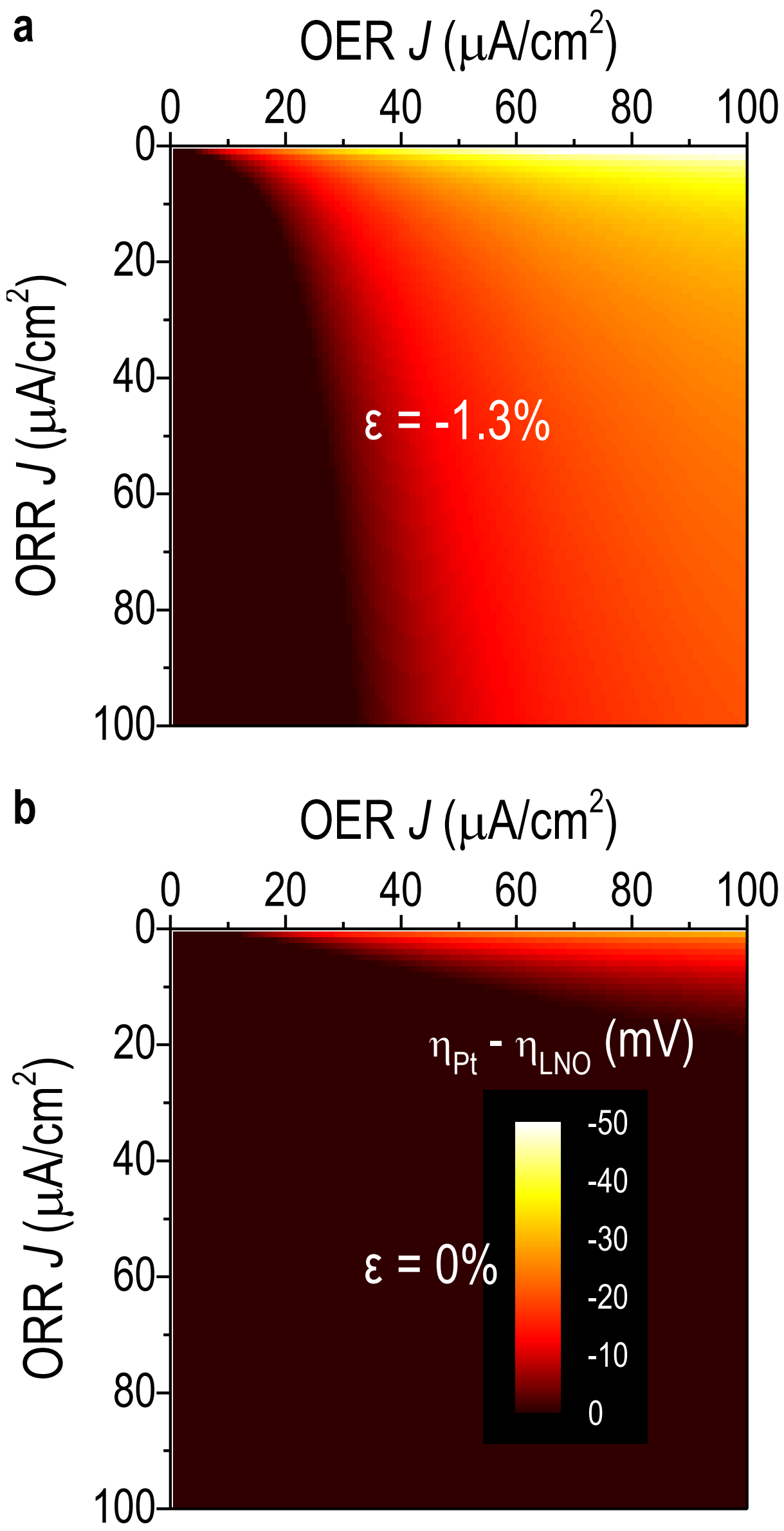

**Figure S7. Comparison of overpotential to Pt for ORR and OER. a,b,** The combined ORR/OER η of LNO compared to that of Pt for ORR and OER at different *J* when at (a) ε = –1.2% and (b) ε = 0. To compare, the total η of Pt is subtracted from that of the film. Negative values of this difference in the bifunctional η between Pt and LNO signify a better bifunctional oxygen activity than Pt. The compressed film has a much larger range of bifunctional *J* where the ORR/OER current is more active than Pt, increasing from the expected high OER *J*/ low OER *J* to a more evenly distributed bifunctional current, as seen in Fig. 1e.

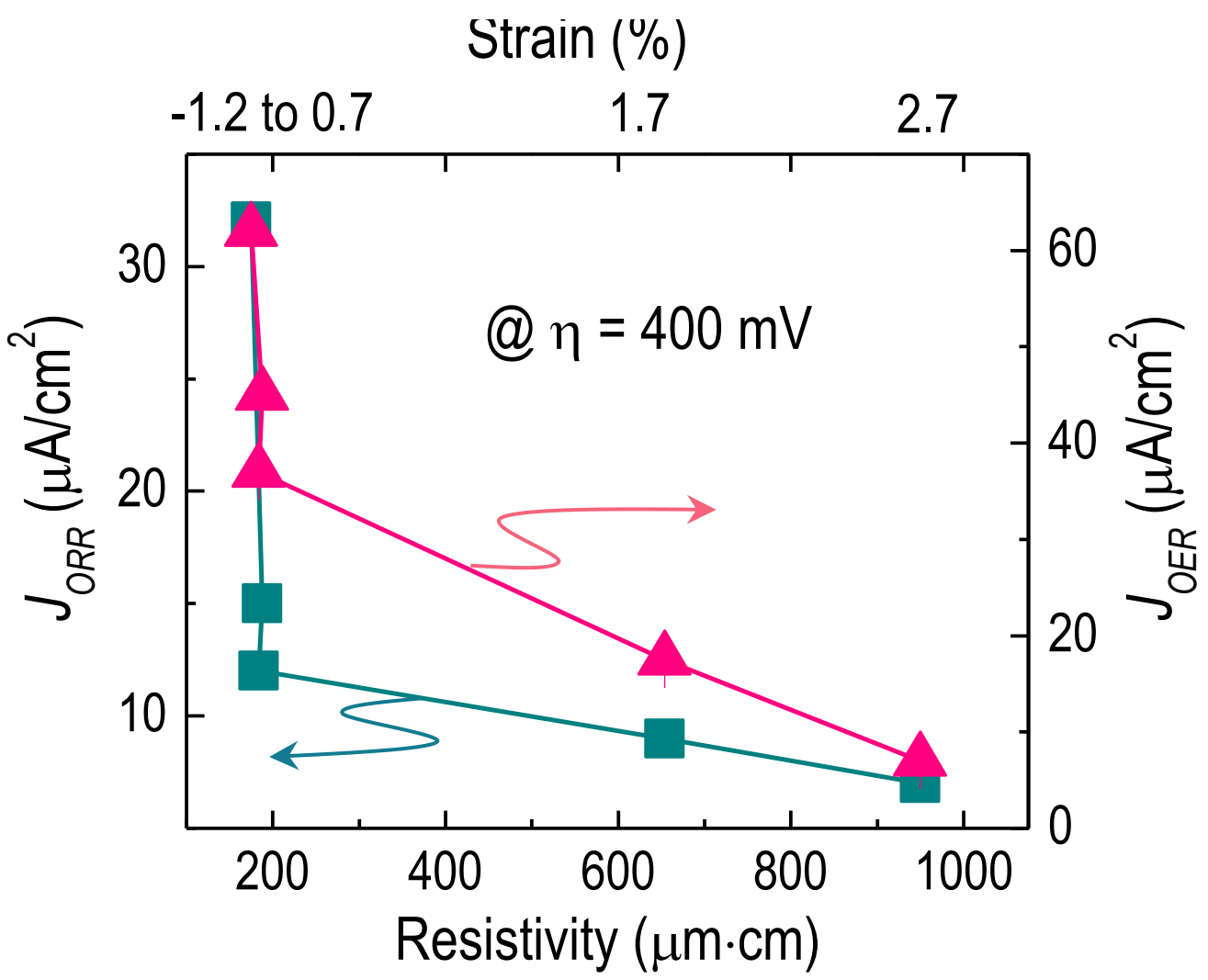

**Figure S8. Activity of strained LNO films as a function of room-temperature resistivity.** By plotting ORR and OER activities (*J* at η = 400 mV) on different substrates as a function of the electrical room-temperature resistivity, rather than strain (as in Fig. 1c), one can see that any correlation between conductivity (inverse of resistivity) and activity breaks down at strains of ε = 0.7% and below. These results argue against changes in electron transport through the material limiting these reactions. As discussed later in Figure S8, resistivity is mainly affected by changes in $NiO_6$ octahedral out-of-plane rotation (tilt) while ORR/OER activity is influenced by changes in Ni-O bond lengths. Near ε = 0.7% and below, tilt remains relatively constant while Ni-O bond length continues to increase, indicating that strain is the root cause of both resistivity and activity changes rather than the former being responsible for the latter.

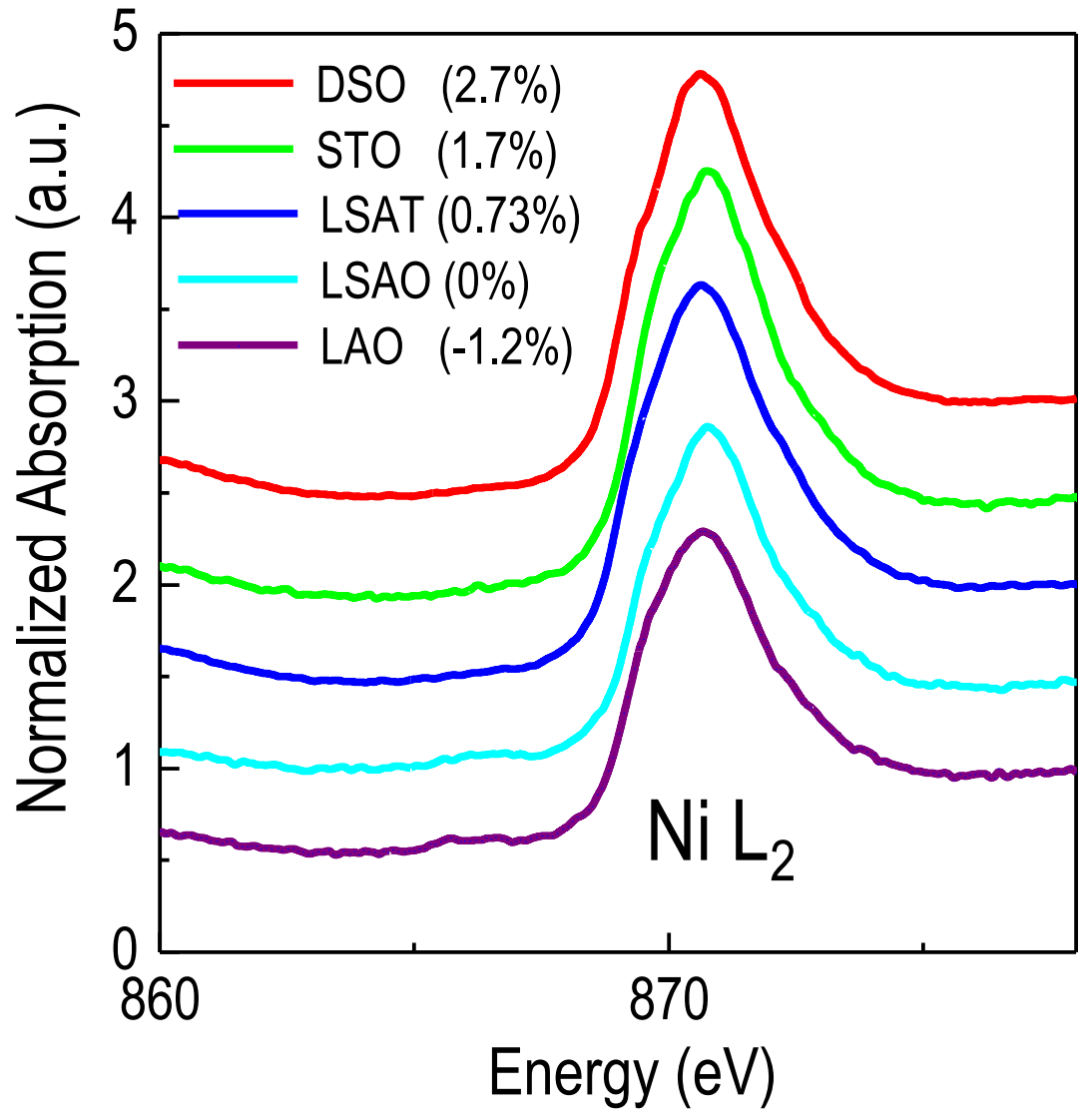

**Figure S9. Determination of Ni oxidation state using X-ray absorption spectroscopy (XAS).** XAS plots of the Ni $L_2$ peak on widely-mismatched substrates. A peak location ~871 eV is indicative of the $Ni^{3+}$ state being predominant.

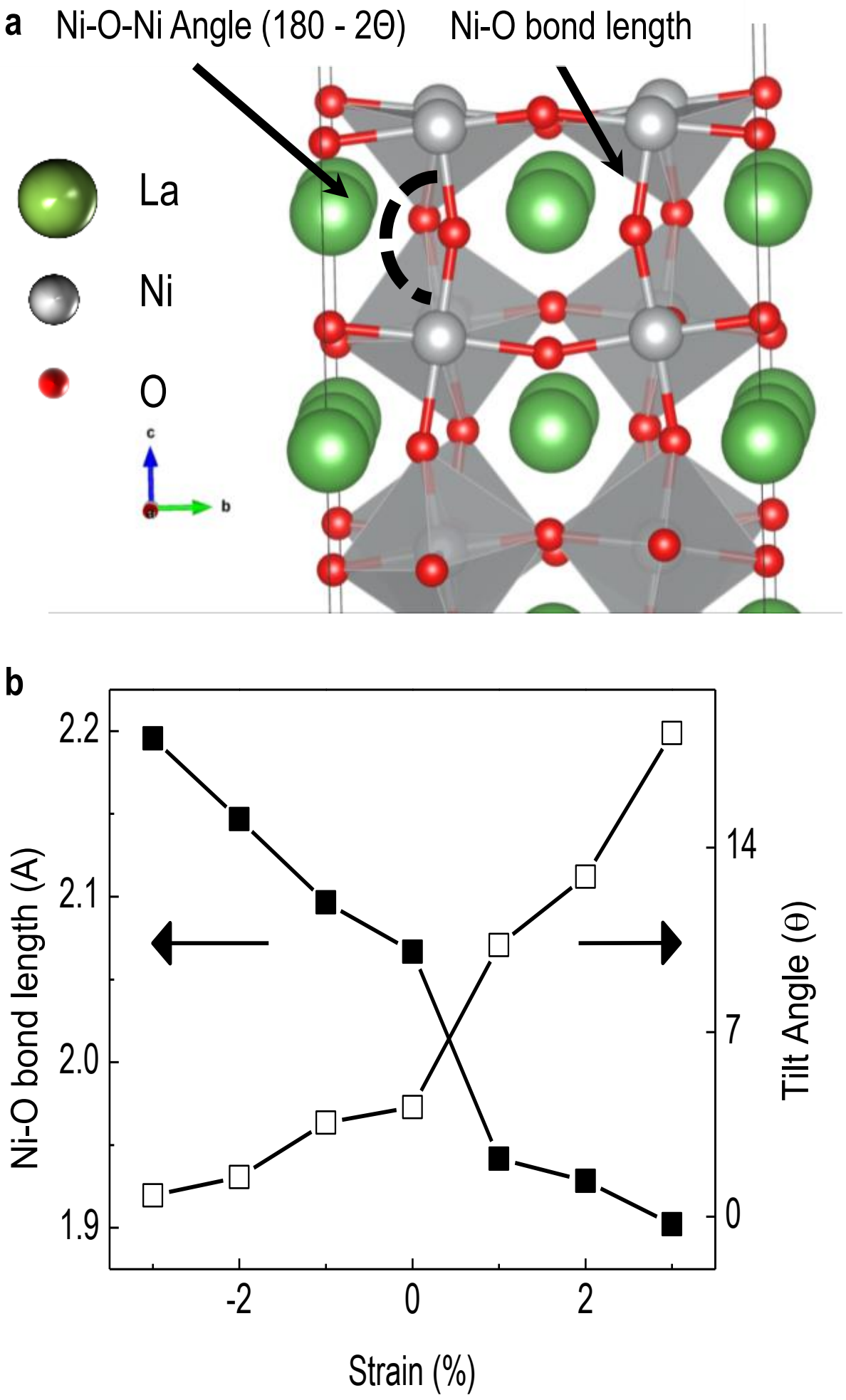

**Figure S10. a,b, DFT modeled tilt and Ni-O bond length near surface as a function of strain.** The rhombohedral structure of LNO accommodates strain via two main mechanisms: (a) rotation and (b) stretching of the $NiO_6$ octahedra. While in-plane rotation is relatively constant, tensile strain results in large increases in out-of-plane rotation (i.e. tilt). This rotation decreases the Ni-O-Ni bonding angle from 164°, resulting in large increases in in-plane resistivity, independently verified via transport measurements (Extended Data Fig. 2). Conversely, sterics from the A-site $La^{3+}$ largely constrain any further reduction of tilt to accommodate compressive strain, favoring the elongation of the out-of-plane apical Ni-O bond length.[5]

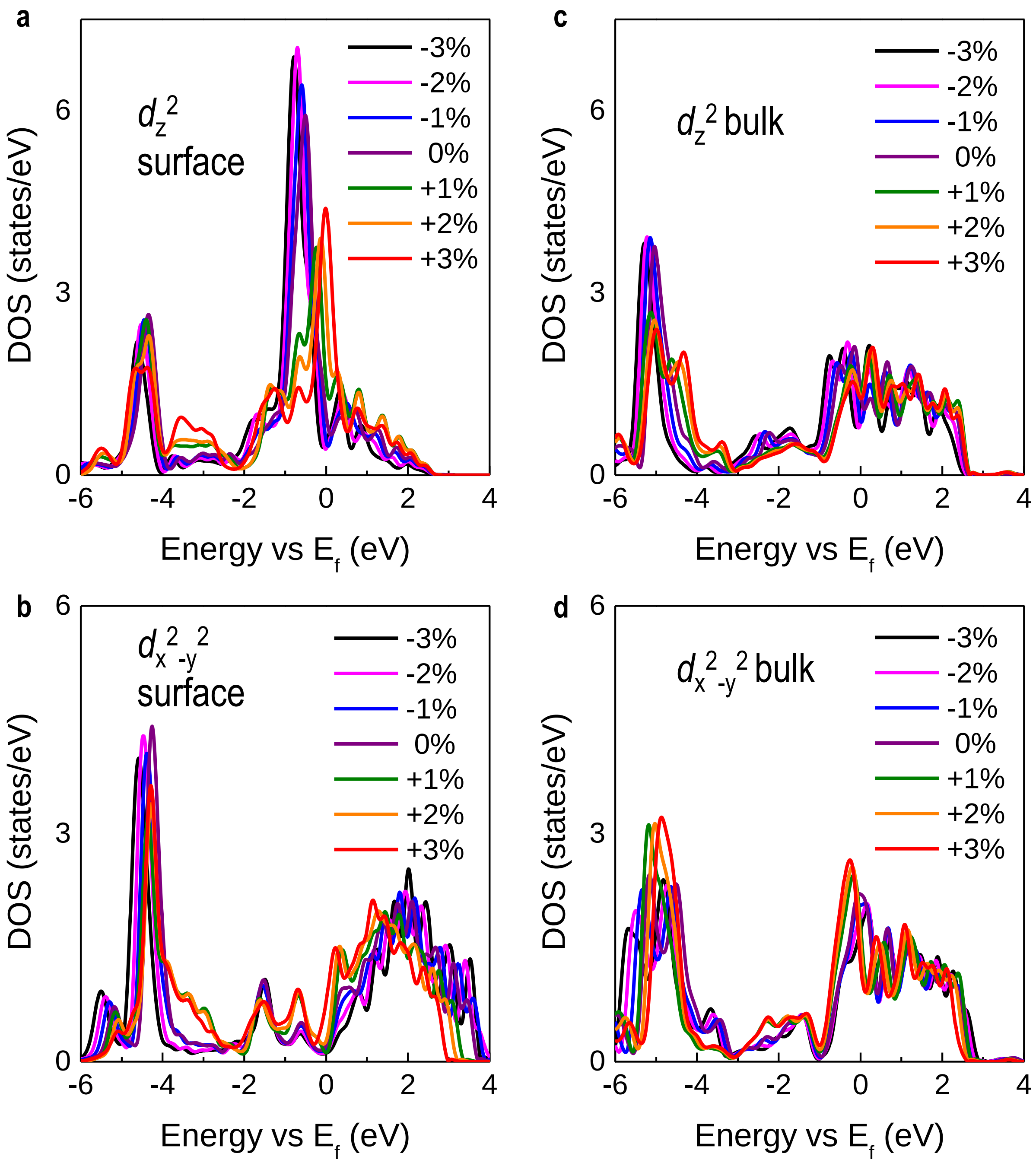

**Figure S11. DFT DOS data in LNO bulk and at surface. a,b,** Density of states (DOS) around $E_F$ (0 eV) for the $d_{z^2}$ orbital on either the (a) first monolayer (surface) or (b) in the bulk, and for the **c,d,** $d_{x^2-y^2}$ orbital on either the (c) first monolayer (surface) or (d) in the bulk. The centroids of each state and electron occupancies are determined via standard integration methods.